Title: Punch Card Programmable Microfluidics

George Korir[1], Manu Prakash*[1]

[1]Department of Bioengineering, Stanford University

318 Campus Drive, Stanford, CA 94305

*Corresponding Author:

Prof. Manu Prakash

Stanford University

318 Campus Drive, Stanford, CA

617-820-4811

manup@stanford.edu

# Title: Punch Card Programmable Microfluidics

Authors: George Korir[1], Manu Prakash[1]*

Affiliations:

[1]Department of Bioengineering, Stanford University

*Correspondence to: manup@stanford.edu

**Abstract**: Small volume fluid handling in single and multiphase microfluidics provides a promising strategy for efficient bio-chemical assays, low-cost point-of-care diagnostics and new approaches to scientific discoveries. However multiple barriers exist towards low-cost field deployment of programmable microfluidics. Incorporating multiple pumps, mixers and discrete valve based control of nanoliter fluids and droplets in an integrated, programmable manner without additional required external components has remained elusive. Combining the idea of punch card programming with arbitrary fluid control, here we describe a self-contained, hand-crank powered, multiplex and robust programmable microfluidic platform. A paper tape encodes information as a series of punched holes. A mechanical reader/actuator reads these paper tapes and correspondingly executes a series of operations onto a microfluidic chip coupled to the platform in a plug-and-play fashion. Enabled by the complexity of codes that can be represented by a series of holes in punched paper tapes, we demonstrate independent control of fifteen on-chip pumps with enhanced mixing, on-off valves and a novel on-demand impact-based droplet generator. We demonstrate robustness of operation by encoding a string of characters representing the word "PUNCHCARD MICROFLUIDICS" using the droplet generator. Multiplexing is demonstrated by implementing an example water quality test utilizing colorimetric assays for pH, ammonia, nitrite and nitrate content in different water samples. With its portable and robust design, low cost and ease-of-use, we envision punch card programmable microfluidics will bring complex control of microfluidic chips into field-based applications in low-resource settings and in the hands of children around the world bringing microfluidics and low-Reynolds number hydrodynamics to everyday classrooms.

**Significance Statement**: The capacity to implement complex robust multiplex assays in resource poor settings devoid of skilled personnel, power sources and supportive infrastructure can lead to a wide range of applications that are currently difficult to execute in medicine, environmental applications and forensics. While multiple approaches have been implemented, success has been limited. Combining microfluidics with programming using punch card tapes, we present an integrated general-purpose platform to address existing challenges. Powered by a hand-crank, our device incorporates a single-layer microfluidic chip in a plug-and-play fashion and is programmed by a paper tape with punched holes as discrete instructions. In addition to the above-mentioned applications, we aspire to enable children to have access to "programmable chemistry kits" in science education settings globally.

The use of microfluidic technology, where small volumes of fluids are manipulated in carrying out miniaturized laboratory assays, has drawn considerable attention owing to inherent advantages that include small reagent consumption, miniaturized reaction volumes and the potential to yield robust and rapid results (1). On-site diagnosis can result in increased rates of disease treatment leading to improvement of the health of people living in low-resource settings (2). There is an especially urgent need for multiplexed tests either to diagnose a disease caused by multiple agents or cases of co-morbidities due to a high disease burden in developing countries (3). Application of microfluidics for robust multiplex diagnostic tests in extremely low-resource settings holds great promise but remains currently unfulfilled due to a variety of challenging factors including absence of electricity, lack of refrigeration for reagent storage, unavailable calibration services for devices over time, challenging operating conditions such as fluctuating temperatures and lack of skilled personnel (3). Therefore a successful implementation in these settings requires surmounting the above-mentioned challenges, using a platform that is completely self-contained and modular in nature, coupled with access to stable reagents that can be easily replenished.

The capability for performing robust, cheap and easy to replicate biological and chemical assays also has new applications in science education settings worldwide. Hands-on introduction of chemical and biological sciences for school children can instill a life-long passion for science (4). Although many current scientists admit to having been inspired by open-ended explorations utilizing chemistry kits widely available several decades ago, safety concerns and expensive reagents have made this exploration currently unavailable. Low-cost self-sufficient microfluidic technologies with enclosed chemicals and small-volume reagent reservoirs, could potentially provide a wide-ranging solution to the problems mentioned above.

With the implementation of pneumatic micro-valves, it is now possible to run thousands of assays in parallel on the same microfluidic chip (5, 6). Although significant progress has been made in development and manufacturing of complex microfluidic chips, current external control systems remain bulky and expensive and often rely on complex equipment such as external syringe or vacuum pumps, gas tanks and computers (3, 7). A few applications for microfluidic devices in educational settings have been explored before, primarily focused on micro-fabrication techniques (8) while others have focused on applying existing platforms to teach principles of fluid dynamics (9). Simpler device fabrication techniques have been developed (10, 11), but controlling complex reactions in small volumes remains out of reach in educational and low-resource settings.

To address some of the challenges brought about by constrains in low-resource settings, several novel approaches have been implemented, including a finger-actuated microfluidic pump device (12) and battery-powered implementation of pneumatic valves using solenoids (13). While promising, such approaches are either limited in range of fluid manipulation or still utilize external solenoid valves that can significantly increase the cost and depend on electrical or battery power-source. Dipsticks and lateral flow assays and in general "paper microfluidics" have found greater success in low-resource point-of-care diagnostics (14-18). Although often low-cost, portable and easy to use, they have limited capacity to run multiplex, complex or a wide range of protocols and are often not as quantitative as traditional microfluidic assays except when paired with specific external readers (19). Furthermore, due to inherent design limitation of capillary flow in a porous medium, paper microfluidics often cannot take advantage of droplet-based assays that are highly sensitive due to further reduction in associated fluid volumes and discrete and isolated nature of trapped fluid samples inside droplets. To address all the challenges listed above, the ideal technology would therefore have the capacity to (a) run complex, programmable, multiplex assays while being self-contained, (b) be capable of handling large fluid volumes in applications where the biological sample has few targeted events, (c) operate with both single phase and multiphase microfluidics and (d) be operable without the need for specialized training or any other external equipment.

Here we present a programmable multiplex microfluidic system based on punch card programming that is hand-crank powered, low-cost, robust, and can run complex biological and chemical protocols with limited chances of human-error. Moreover, our system is rugged, portable (weighing approximately 100 grams and measuring approximately 2 inches in length, 1.5 inches wide and 1 inch high) and is self-contained. The system does not require any external pumps or other supportive equipment to run. Multiple protocols can be run in parallel (multiple assays on the same sample or single assay on multiple samples), manipulating fluids arbitrarily with nanoliter volume precision. Because the program is encoded in punch card tape, the protocols can be easily shared like baseball cards to repeat or modify existing assays.

Our current implementation is inspired by punch card programming as historically applied in a wide range of applications beginning with control of textile looms (20), early computing (21), music replay (22) and data processing. Other implementations of technology that used punch cards include the control of radio telescopes (23), and voting machines (24). Punch cards enable the use of a single actuating platform to execute multiple programs resulting in implementation of complex instruction sets that yield radically different outcomes by simply switching the punch card tape. Such a platform offers the flexibility of achieving multiple results without the need for redesigning the system for new tasks. We have harnessed this approach and implemented it to manipulate fluids in a low-cost platform.

**Mode of operation**
Our system is comprised of paper-based punch card tape a polydimethylsiloxane (PDMS) based single layer microfluidic chip and a mechanical reader/actuator that couples fluidic channels to paper tape (Fig. 1 *A-F*). The punch card stock paper tape serves as the program encoding the protocol to be run. Our current implementation consists of a series of 15 parallel lines on which holes can be punched to actuate 15 independent channels (Fig. 1*F*). Thus the device has a bandwidth of 15 independent bits that can be set at any given time. Our platform has three key components required in general purpose microfluidic processors: embedded flow-controlled micro-pumps, ON/OFF valves and a novel impact based droplet generator (Fig. 1*B-D*). The mechanical reader/actuator powered by a hand crank (Fig. 1*A*, inset) interfaces with the punch card tape to uniquely read the punched code and execute pumping, valving and droplet generation in a microfluidic chip. Our modular design enables microfluidic chips to be inserted and removed from the device in a plug-and-play fashion.

Protocols are encoded, stored and executed using card-stock paper tapes (width 41 mm). Presence or absence of a series of small holes (5mm diameter) punched on paper ticker tape provides a simple 15-bit encoding scheme (read in parallel). The spacing between two adjacent tracks on paper was set to 2 mm, based on the geometry of the current mechanical reader/actuator. This width helps avoid inadvertent actuation of neighboring channels. The spacing (along the tape, parallel to the actuator gear train) between holes sequentially passing through the device determines the relationship between the timing of actuation of corresponding channels. Based on the design of the actuator gear teeth (see Fig. *S1* for details of the mechanical design), it was determined that for each gear tooth, holes should be positioned at least 1 cm apart along the tape for unique actuation for each hole punched. No limits are placed on the length of the program (governed by the length of the punch card tape). For specific applications, the same program (punched tape) can also be run in a circular configuration for continuous and repeated implementation.

For the current reader/actuator implementation, we exploit the mechanism of a Kikkerland Music Box$^{TM}$ - a toy readily available in the market. Although many such mechanical music toys exist, we utilize the open architecture in the Kikkerland$^{TM}$ setup for quick prototyping. A completely 3D printed version of the reader/actuator was also implemented (Fig. *S1*, inset) for ease of distribution and as an initial step in the rapid development of subsequent generations of our device. The reader/actuator consists of a gear train powered by a hand-crank coupled to a rotating rod that reels in the punch card tape with an approximate gear ratio of 1:6. Actuation frequency can be made independent of hand-crank frequency by implementing a spring based storage mechanism that provides a constant rotational output independent of the hand crank rotation speed. To initialize the device, punch card tape is inserted in a slot comprised of thin metal sheets that act as guides toward two counter-rotating rods that are coupled to a series of gears. The plastic encasing of one of the rods provides additional friction enabling the turning rods to effectively reel in the punch card tape as the hand-crank is turned. The driven rod coupled to the gear train also consists of 15 concentric metal discs that have four asymmetric teeth spaced 90 degrees apart from each other (Fig. 1 *B-D*). Plastic spacers between metal discs result in a 2 mm spacing between adjacent discs with gear teeth. The discs have the capacity to move independently and are only engaged when a punched hole appears in the paper tape. As the punch card tape is reeled in, the gear tooth that is closest to the hole eventually gets caught up in the hole and gets pushed in the direction of the actuation of punch card tape (+Y-axis). This action results in a rotation of the disc and therefore the other three gear teeth on the same disc also rotate in concert. For implementation of valves and droplet-generators, a secondary cantilever based array of pins is also coupled to the rotating gear train, enabling a single hole to actuate a vertical pin (Z-axis) to move up and down (see Supplementary Movie *S1*).

A microfluidic chip is coupled to the device in a plug-and-play fashion, held in place only by friction. Our chips consist of patterned single layer microfluidic channels on a thin membrane of PDMS (thickness 500 $\pm$50 $\mu$m) and couples directly to the gear train from the reader/actuator. The microfluidic channels are aligned to gear teeth for fluidic pumping or valving with each actuation instance. We mold single layer microfluidic channels using standard soft-lithography with channel height of 50 $\mu$m and width of 200 $\mu$m. This thin film is mounted on a thicker slab of PDMS that provides mechanical support, inlet and outlet channels and means to couple the device to the reader/actuator. Here, we demonstrate pumping, enhanced mixing, valving and on-demand droplet generation.

## Pumping

Integrated microfluidic pumps remove the dependence of a fluid-handling platform on external, bulky syringe pumps or pressure sources. In the past, several implementations for microfluidic pumps have utilized the deformable nature of PDMS channels including pneumatic quake-valve pumps (5), braille display based devices (25), finger-actuated flows (12), passive capillary pressure pumps (26) and external motors driving pins that actuate fluidic cavities (27). While effective, they often require external components (solenoid valves) and/or pressure sources and often do not provide independent multiplexed control.

In our current platform, we have implemented 15 integrated and independently controlled micro-pumps where net output flow-rate can be controlled as a function of actuation frequency and actuation height (Fig. 1 *B and F*). The actuation frequency is dependent on the total number of punched holes on the paper tape per unit distance and the angular velocity of the gears in play. The actuation height is determined by how the microchip interfaces with the gear teeth. It is crucial to note that our implementation requires no external component (pressure lines or even any external tubing or connectors) to run or operate these integrated pumps. The only power source for the device is a simple hand-crank. The pumps can be further operated with or without integrated valves, as described in later sections.

The basic principle of the pump is illustrated in Figure 1B, where a rotating disc couples a gear-tooth to a completely collapsible microfluidic channel. As the disc rotates, the squeeze contact point linearly translates the collapse position along the channel (Fig. 2*A*). The collapsed channel can be easily imaged from above and the indentation caused by the collapse covers the entire channel. The width of the channels (200 µm) was further chosen to be less than the gear teeth width at the actuation area (~480 µm wide) to ensure that the channel collapse is complete with each pumping cycle. The gear tooth engages with a given channel once for a single hole. Thus one of the variables that controls the flow-rate of the pump is the total number of punched holes encountered per unit time.

The vertical position of the microfluidic chip with respect to the reader/actuator determines the degree of collapse of the microfluidic channel. Thus we can quantify the fluid flow in the channels with two parameters: height of the chip ($h$) above the actuator and angular velocity of the gear disc under rotation ($\omega$). To characterize the pumping, fluid flow was imaged using red fluorescent microspheres (size 2 µm) (Fig. 2*B*). Individual trajectories of the beads, depicted as kymographs reveal the oscillatory dynamics of the fluid flow as a function of $h$ and $\omega$ (Fig. 2*C*). For the purpose of data collection, precise angular velocity ($\omega$) was implemented using a simple motor driving the hand-crank. The oscillatory dynamics arise from the complete collapse and ensuing recoil of the channel as a result of the elasticity of PDMS, leading to a forward and backward flow respectively. The asymmetry in forward and back flow is introduced due to preferential movement of the channel collapse point in the forward direction. The net effect of the back flow is felt less with increasing actuation height (Fig. *2E*). For angular velocity ($\omega$) of 1.5 rotations per second and displacement height $h$=450 µm, a net flow rate of ~30 nL/s can be easily accomplished. Below the critical value $h$=50 µm, no net forward flow is observed in the channels due to insufficient coupling of the rotating gear disc with the PDMS channel. With increasing actuation height, the arc across which total channel collapse occurs with each actuation increases leading to increased fluid flow with each rotational stroke (Fig. 2*E*). The net fluid flow increases in concert. The microfluidic chip eventually fails when the pump gear tooth tears through the PDMS channel through friction with time or is unable to overcome the force exerted on it by the microfluidic chip. An experiment to determine wear of the channels was run at an angular velocity of 1.5 rotations per second and actuation height leading to a net flow rate of 10 nL/s. After 100 hours of continuous operation, the channels were still functional although worn down from the friction between the gear teeth and PDMS. Scotch™ tape applied on the PDMS at the point where it interfaces with the gear teeth eliminated the wear on the PDMS due to friction for an experiment that was also run for 100 hours at an angular velocity of 1.5 rotations per second and actuation height leading to a net pumping flow rate 30 nL/s. To demonstrate multiplex operation, six independent pumps were operated with a zig-zag pattern of punched holes in a paper tape (Fig. 2*D*), resulting in contiguous operation of multiple micro pumps.

**Enhanced fluid mixing**
Fluid mixing is a significant challenge in integrated microfluidic devices due to lack of fluid inertia at low Reynolds numbers. In the past, complex micro structures such as herringbone geometry (23) have been utilized to implement mixing in single-phase flow by extending the interfacial boundary between two miscible fluids, thus increasing the effectiveness of diffusion dominated mixing. Here, we present a simple strategy for fluid mixing controlled by a zig-zag actuation pattern encoded on a punch card tape.

By exploiting the pulsatile nature of our pumps with precise control of time-varying actuation of adjacent channels, we implement a simple mixing strategy in our devices (Fig. 3). As a demonstration, we mixed six different water-based fluids simultaneously in a single output channel. The fluid was pumped through the device using the punch card tape at a flow rate of approximately 100 nL/s with an effective Re number of 0.2. Even at such low Reynolds numbers, efficient mixing was achieved within 16 mm downstream of the inlet channels. Since all pumps can be independently controlled by 15-bit punch card tape, the diffusion boundary between nearby fluid streams (for instance, red and green streamlines) can be folded significantly by offsetting the moment of pulse generation. This is achieved using punch card tape with a zig-zag pattern of pumping (Fig. 3*A*, right inset). Although all channels have the same average flow rate (due to similar number of holes per unit distance), the time-varying flow rate induces interfacial boundary folds enhancing diffusion based mixing (as depicted in the time series in Fig. 3). Various other punch tape patterns can be utilized to explore efficiency of mixing by folding flow lines.

To confirm that the folding only arises due to offset and pulsatile nature of our punch card controlled integrated micro pumps, we ran an equivalent experiment using a traditional syringe pump replicating the exact flow rate, as generated by our integrated micro-pumps. As can be seen, the streamlines do not mix and the six coored fluids remain effectively separated (Fig. 3, left inset). We further quantify the extent of mixing in the six fluid streamlines from the six colored fluids pumped using a mean shift clustering algorithm implemented on images taken along different points (white boxes marked a,b,c,d) of the fluidic channel for a test case of six fluid lines (labeled with food color in water). In this nonparametric clustering technique, prior knowledge of the number of clusters is not required (28). Pattern recognition is based on the mean shift approach (29) applied to the analysis of individual images (30). The analysis entails shifting each data point to the average of data points around it in an iterative fashion until there is convergence, and then clustering the data points around the closest convergence point of a particular neighborhood radius. For each chosen section, an image was taken and the red, green and blue values of each pixel were taken as data points to be clustered. The data was then processed using the mean shift algorithm and the pixel values clustered for each image using the same bandwidth (radius of the neighborhood). Prior to mixing, six clusters corresponding to the six fluid colors were identified (Figure 3B) and as samples were analyzed along the fluid channel, the number finally reduced to two clusters within a distance of 14 mm along the outlet channel (Fig. 3 *E*).

**Valves**
Successful and arbitrary manipulation of fluids to run a wide range of chemical and biological assays requires the use of micro-valves to enable programmable spatiotemporal control of fluid flow. Combined with integrated pumps, valves can easily facilitate arbitrarily complex flow control strategies, as have been previously demonstrated (5, 6) in multi-layer microfluidic valve structures. Numerous large-scale integration architectures based on valving schemes have also been described, including precise design rules for operation (31). Most implementations of dynamic programmable valves involve external electrical solenoids, external pressure sources and expensive electronic control. The need for external control systems often limits the impact that valve based microfluidics could have in resource-poor settings.

We successfully implemented multiple, independently controlled, punch card programmable micro-valves in the current device (Fig. 4). A cantilever mechanism was implemented coupled to gear discs on all independent channels (Fig. 4 *B*). The valving mechanism comprises of independent pins (~400 μm diameter) projecting perpendicularly (along the Z-axis) from a series of cantilevers that are attached to the base plate of the reader/actuator (Fig. 4). In the default setting, the valve pins are passively aligned to the microfluidic channels pushing against the PDMS and thus collapsing the channel at the point of contact (OFF setting, Fig. 4 *D* and *E*). During actuation, the gear train teeth pluck the cantilever beams thus opening the channel due to the downward and forward motion of the valve pin. The channel is closed passively due to the elastic deflection of the cantilever (Fig. 4 *C* and *D*). A single ON/OFF cycle of a valve can be implemented in approximately 0.5 s (Fig. 4*D*). The implementation described above needs to be actuated to be in the OFF state (ON being the default state) to let fluid flow. A reverse valve that is OFF in the default mode can be implemented by simply reversing the leverage point of the cantilever where a downward pull pushes the pin upwards and vice versa.

As a demonstration, we further implemented integrated valving and pumping in our device. The teeth on a gear disc (Fig. *S3*) coupled for pumping were positioned such that as one interfaced with the punched hole, the ensuing rotation led to the corresponding rotation of the gear tooth that actuated the valve while another gear teeth pumped the fluid simultaneously. The coupled action allowed the valve for the micro-pump to be open only when the fluid was being pumped and remained closed as soon as this action was complete. Because 15 independent punch card tape channels are available on the current implementation, the same number of independent valves can be operated simultaneously. Changing the dimensions of the reader/actuator will easily enable a greater number (~100) of independent valves.

**On-demand droplet generation**
Droplet microfluidics has been used to implement multiple highly sensitive assays (32-34). Current on-demand droplet generation (35) and control (36-38) in microfluidic channels for complex assays has often been demonstrated using complex external controllers. Here we demonstrate a new on-demand droplet generator programmed using our punch card ticker tape by utilizing impact dynamics on a soft substrate to enable controlled single droplet dispensing.

The droplet generator presented here relies on the impact dynamics of a small pin-head on a compliant microfluidic channel to generate a small fluid jet which destabilizes very quickly (in 2 ms) resulting in a single mono-disperse droplet dispensed in surrounding carrier fluid (See Supplementary Movie S1 and Fig. 5*A*,*B* and *D*). Unlike dynamic droplet generators that rely on co-axial continuous mechanisms (39), resulting in a continuous stream of droplets, only a transient pressure pulse (generated by pin impact triggered from a punched hole) is utilized in our system.

The impact droplet generation mechanism was implemented using the same actuation pins mounted on cantilever beams used for valving (Fig. 4*B*). The stiffness of the cantilever beam can be further tuned to regulate the time dynamics and remove any associated ringing effects upon impact. The impact pin height was tuned so as to only allow the cantilever mounted vertical pins to interface with the channel while the pumping gear teeth are too low to interface with the PDMS micro-channels. The impact pins had a default "ON" position that led to complete channel collapse sealing the micro-channel. A flow-focusing geometry with a capillary valve was used (Fig. 5*A*, inset) for on-demand droplet generation. The carrier fluid (filled with mineral oil) channels have a small flow rate (~0.5ml/min) to move the droplets along the channel and play no role in droplet generation.

For experiments reported in Figure 5 (and Supplementary Movie S3), mineral oil was used as the carrier fluid with water (Tween 20, 2% v/v). One-to-one correspondence is achieved between a single hole in punched card tape, impact of the corresponding pin and a single droplet generation in the corresponding micro-channel (Fig. *5C*, Movie *S3*). Although the punch card holes only arrive every few seconds, the actual drop formation process is nearly instantaneous (~50 ms). Droplet formation was filmed with a high-speed camera (Fig. 5*D*, Movie *S4*) with the time series depicting the unusual shape dynamics of the jet and its subsequent collapse. Our configuration dispenses water droplets in higher viscosity mineral oil (viscosity, ~15 cP at room temperature), resulting in a set of jet dynamics that is strikingly different from dispensing water in air (41).

To further demonstrate the robustness of our system, we tested our on-demand impact droplet generation method to produce single droplets (volume ~2.5 nL) driven by a long punch card code, thus establishing one-to-one mapping. We implemented a binary code consisting of 22 characters to write out "PUNCHCARD MICROFLUIDICS" using droplets generated by the impact droplet generation method (Fig 6*A*). Five bits were used to encode each letter of the alphabet, and seven bits (all of which were zeros) were used to encode the space between words. The coding system followed the binary system such that the letter A was represented by "00001", B = "00010", C="00011"… and so on. The code was programmed into the punch card tape by punching holes only whenever the high signal ("1") was required. The "0" signal was represented by absence of the punched holes, leading to no droplets being generated. In the results displayed on Figure 6A, there was an error on the letter "O" for the word "MICROFLUIDICS" that arose at the transition point between two punch card tapes joined together by Scotch™ tape that obscured a punched hole. Each high signal led to the generation of a single 2.5 nL droplet throughout the operation of the device.

**Multiplex assay**
We have presented a platform for general-purpose control of microfluidic systems using punch card ticker tapes capable of multiplex assays. To demonstrate the ease of use of our platform, a simple qualitative colorimetric assay to test for pH and the presence of ammonia, nitrates and nitrites in three water samples was implemented (Fig. 6*B*). The samples consisted of deionized water, pond water from Palo Alto, CA and filtered seawater from Monterey Bay, CA. The reagents used were from the API Colorimetric Water Test Kit (Aquarium Fish care, see methods section for implementation details). Figure 6*B* reveals the results comparing a macro scale implementation to the output from our microfluidic outlet channels. Based on this quick colorimetric readout, we were able to qualitatively demonstrate the differences in pH and presence of ammonia, nitrates and nitrites, across the three water samples.

**Discussion**
In our current paper, we have demonstrated integrated microfluidic pumps, valves, mixers and on-demand droplet generators all of which are controlled by a paper punch card tape only. The current device is limited to 15 independent channels based on the width of the punch card tape being read. This is not a fundamental limitation and can be dramatically increased by simply choosing higher density line spacing for gear discs or a wider punch card tape. Further increase in the number of control valves can be achieved by multiplexing individual valves using blind channels filled with ionic fluids (25). One current limitation of our implementation includes wear and tear of the microfluidic chip surface over a long-time usage for applications that require processing of large fluid volumes or long protocols. The wear and tear can simply be mitigated through the use of Scotch™ tape applied on the interface between the gear teeth and the PDMS. An experiment ran at an actuation height that would result in fluid pumping at 30 nL/s demonstrated no wear on the PDMS chip after four days of continuous running when Scotch™ tape was used at the interface between the gear teeth and the PDMS at a lower actuation height. A parallel experiment for a chip actuated without the Scotch™ tape at a lower actuation height (for a pumping rate of 10nL/sec) led to wear through friction. Another strategy would be through using material coatings on the chip and the metallic gear pin to reduce friction, or the fabrication of gear teeth using materials that lead to reduced friction upon contact with the PDMS.

A reliable supply chain of reagents required for biological and chemical assays is difficult to maintain in remote resource-constrained settings. Further, the manipulation and storage of reagents currently use containers that are often expensive can be difficult to dispose of (40). To address some of these challenges, a recent paper Bwambok et al (41) demonstrated the use of bubble wrap pouches to store reagents and samples in liquid form and performing assays on them. In our approach we envision addressing this problem by embedding dry reagents (lyophilized) in both the punch card tape and/or the microfluidic devices instead of liquid samples for increased stability and shelf life. This would provide a single means to both provide all the necessary reagents, tools and the protocol to run a given assay for optimal delivery. As next steps, we will encapsulate reagents in the paper punch card medium that will allow for rupture on one side and will interface with the microfluidic chip on the other. An example of such a packaging is medical-tablet blister packs that are also compatible with roll-to-roll manufacturing. Such a configuration combines the best of both worlds, by utilizing dry reagents (for instance through lyophilization or the encapsulation using pullulan extract (42)) with the capability to run non-linear, multiplex and complex assay and protocols (adapted from multi-layer microfluidic valves). This functionality will minimize reagents used, enable the running of multiple complex assays with minimal training and no external handling of expensive reagents while also making it possible to easily ship and replenish reagents. The final result could also be deposited back on the paper tape thus providing a record of the medical test performed for later analysis or reference.

Here we have demonstrated design principles behind a general-purpose punch card microfluidic device. In the future, we will design and print readers/actuators with more features specific to the diagnostic tests being conducted and further increase the channel bandwidth (from 15 to 30) by simply increasing the density of punch card reading channels. We will further implement several example assays for educational settings and test them with children as a user base for further feedback. We will develop assays for medical diagnoses and environmental applications.

**Conclusion**
Combining punch card tapes with microfluidics provides a novel, scalable and inexpensive yet robust means to enable multiplex, general-purpose control of complex microfluidic chips. The platform is simple enough to use by both untrained health workers and young children. The assay is run by simply inserting an appropriate punch card tape with a corresponding microfluidic chip with pre-deposited reagents, making it easy to use and replicate anywhere. Untrained users anywhere in the world can replicate protocols encoded in the ticker tape. We believe that combining the capability to program general purpose microfluidic systems using punch card tapes and utilizing a hand-crank based power source can bring a broad range of capabilities from the lab to real-world field conditions. This makes future diagnostic instruments built on the presented platform truly portable. We envision that with our approach, we will also bring hands-on chemistry and biological manipulation into the hands of a wide range of new users in educational and resource-poor settings, opening up citizen science for curiosity driven explorations.

## Materials and Methods

*Microfluidic chip design and fabrication*

We fabricated our microfluidic chips using standard PDMS photolithography techniques (43). The device was composed of two PDMS layers. The channel features were fabricated to have a height of 50 $\mu$m and a width of 200 $\mu$m. The microfluidic channel geometry included a flow focusing design for droplet fluidic experiments. The channels were spaced 2 mm apart so that they could be aligned to the actuators. In addition, the number of actuated channels was no more than 15, the maximum limit of the current reader/actuator. Please see the supplementary materials for additional details

*Actuator and punch card tape*

For the current implementation of the actuator, we used a Kikkerland Music Box$^{TM}$ and modified it as required, including new gear teeth profiles and associated cantilever beams with vertical pins for valving and impact droplet generation. Punch card tapes (41 mm wide) were made out of simple card stock paper with guidelines drawn for punching holes. No limit was imposed on the length of the punch card tape. For current experiments, cantilever based valve pins were designed using AutoDesk Inventor and fabricated by 3D printing using a multi-jet modeling ProJet 3500 HD printer (Figure 4B). This component was then fastened onto the base plate of the punch card driver in a way that allowed for programmable actuation of the valving mechanism.

*Reagents Used*

Mineral oil (Fisher Scientific Catalog No. 8042-47-5) was used as the carrier fluid in the on-demand droplet generation experiments. The fluids used in the single-phase experiments and for the droplets immersed in the carrier fluid were made using food coloring (Wilton 8 Icing colors set) mixed with deionized water at a 1:10 v/v ratio. 2%v/v Tween 20 was added as a surfactant for all experiments.

*Water quality test*

A colorimetric assay for testing various compounds in water was implemented using the API Water Test Kit (Aquarium Fish Care Catalog No. 347). Filtered pond water, sea water and deionized water were used as samples to be tested. The punch card code was designed to match the volumetric ratio between the reagents and the water sample to be tested according to the manufacturer's test kit protocol. The code was a one-to-one mapping of punched hole to reagent fluid volume introduced into the sample water line. Through valving, the reagent was introduced into the sample chamber only upon actuation. A multiplex assay consisting of five different reagent lines was implemented to test the water samples simultaneously. The colorimetric readings were done at the channel outlet where the fluid column along the light path was longest due to the combination of both the thick PDMS slab and the thin channel layer. Images presented were taken using a Canon EOS Rebel t4i DSLR camera. See supplementary materials for additional details.

*Flow Velocity Characterization Experiments - Pumping*

To characterize the pumping capabilities of our device, red fluorescent 2 µm monodisperse polysterene microbeads (Polysciences Catalog No. 18660-5) were added to deionized water at a concentration of approximately 10,000 microspheres per microliter. The solution was then pumped with our device through a microfluidic channel measuring 200 µm wide, 50 µm deep and 2.5 cm long (Fig. 2*B*). The punch card tape used to drive the actuator had holes punched 1 cm apart along the entire length of the tape. The ends of the punch card tape were joined together using Scotch™ tape for a continuous loop whenever the device was activated. A motor (Maxon DC motor Catalog No. 166789, Maxon Gear Catalog No. 218418) was attached to the hand-crank handle for constant angular rotation. The microchip was lowered onto the gear teeth in increments of 12.5 µm from when the microchip just made contact with the gear teeth to 450 µm. Four different angular rotations (36, 60, 72 and 90 rpm) were used for each actuation height tested. By varying the angular rotations, the actuation frequency was thus varied to investigate the pumping rate as a function of hand-crank rotation frequency. See supplementary materials for additional details.

# References

1. Whitesides GM (2006) The origins and the future of microfluidics. *Nature*, 442:368-373.
2. Hay Burgess DC, Wasserman J, & Dahl CA (2006) Global health diagnostics. *Nature* 444 Suppl 1:1-2.
3. Yager P, *et al.* (2006) Microfluidic diagnostic technologies for global public health. in *Nature*, 442:412-418.
4. DeJarnette NK (2012) America's Children: Providing Early Exposure to STEM (Science, Technology, Engineering and Math) Initiatives. in *Education* 133(1):77-84
5. Unger MA, Chou HP, Thorsen T, Scherer A, & Quake SR (2000) Monolithic microfabricated valves and pumps by multilayer soft lithography. in *Science* 288(5463):113-116.
6. Araci IE & Quake SR (2012) Microfluidic very large scale integration (mVLSI) with integrated micromechanical valves. in *Lab on a chip* (The Royal Society of Chemistry) 12(16):2803-2806.
7. Chin CD, Linder V, & Sia SK (2007) Lab-on-a-chip devices for global health: Past studies and future opportunities in *Lab on a chip* (Royal Society of Chemistry), 7(1):41-57.
8. Berkowski KL, Plunkett KN, & Yu Q (2005) Introduction to photolithography: Preparation of microscale polymer silhouettes. in *Journal of chemical education* 82(9):1365
9. Hemling M, Crooks JA, & Oliver PM (2013) Microfluidics for High School Chemistry Students. in *Journal of chemical education* 91(1):112-115
10. Yang C, Ouellet E, & Lagally ET (2010) Using inexpensive Jell-O chips for hands-on microfluidics education. in *Analytical chemistry*. 82(13):5408-5414
11. Yuen PK & Goral VN (2010) Low-cost rapid prototyping of flexible microfluidic devices using a desktop digital craft cutter in *Lab on a chip* (The Royal Society of Chemistry) 10(3):384-387.
12. Li W, *et al.* (2012) Squeeze-chip: a finger-controlled microfluidic flow network device and its application to biochemical assays. in *Lab on a chip* (The Royal Society of Chemistry) 12(9):1587-1590.
13. Addae-Mensah KA, Cheung YK, Fekete V, Rendely MS, & Sia SK (2010) Actuation of elastomeric microvalves in point-of-care settings using handheld, battery-powered instrumentation. in *Lab on a chip* (The Royal Society of Chemistry) 10(12):1618-1622.
14. Cheng CM, *et al.* (2010) Paper-based ELISA. *Angewandte Chemie* 49(28):4771-4774.
15. Martinez AW, Phillips ST, Butte MJ, & Whitesides GM (2007) Patterned paper as a platform for inexpensive, low-volume, portable bioassays. *Angewandte Chemie* 46(8):1318-1320.
16. Martinez AW, *et al.* (2008) Simple telemedicine for developing regions: camera phones and paper-based microfluidic devices for real-time, off-site diagnosis. *Analytical chemistry* 80(10):3699-3707.
17. Pollock NR, *et al.* (2013) Field evaluation of a prototype paper-based point-of-care fingerstick transaminase test. *PloS one* 8(9):e75616.
18. Pollock NR, *et al.* (2012) A paper-based multiplexed transaminase test for low-cost, point-of-care liver function testing. *Science translational medicine* 4(152):152ra129.
19. Yetisen AK, Akram MS, & Lowe CR (2013) Paper-based microfluidic point-of-care diagnostic devices. in *Lab on a chip* (The Royal Society of Chemistry) 13(12):2210-2251.
20. Hobsbawm E (2010) Age of Revolution 1789-1848.
21. Cortada JW (2000) Before the Computer: IBM, NCR, Burroughs, and Remington Rand and the Industry They Created, 1865-1956.
22. Reblitz AA & Bowers QD (1981) Treasures of Mechanical Music
23. Hodgson JH (Geological Survey of Canada, Open File 1945).
24. United States. (2002) *Help America Vote Act of 2002* (U.S. G.P.O. : Supt. of Docs., U.S. G.P.O., distributor, Washington, D.C.) p 65 p.
25. Gu W, Zhu X, Futai N, Cho BS, & Takayama S (2004) Computerized microfluidic cell culture using elastomeric channels and Braille displays. *Proceedings of the National Academy of Sciences of the United States of America* 101(45):15861-15866.
26. Walker G & Beebe DJ (2002) A passive pumping method for microfluidic devices. *Lab on a chip* 2(3):131-134.
27. Chen D, *et al.* (2010) An integrated, self-contained microfluidic cassette for isolation, amplification, and detection of nucleic acids. *Biomedical microdevices* 12(4):705-719.
28. Stroock AD, *et al.* (2002) Chaotic mixer for microchannels. *Science* 295(5555):647-651.
29. Comaniciu D & Meer P (2002) Mean shift: a robust approach toward feature space analysis. in *Pattern Analysis and Machine Intelligence, IEEE Transactions on* (IEEE) 24(5):603-619.
30. Fukunaga K & Hostetler L (1975) The estimation of the gradient of a density function, with applications in pattern recognition. in *Information Theory, IEEE Transactions on* (IEEE) 21(1):32-40.
31. Cheng Y (1995) Mean shift, mode seeking, and clustering. in *Pattern Analysis and Machine Intelligence, IEEE Transactions on* (IEEE), 17(8):790-799.
32. Melin J & Quake SR (2007) Microfluidic large-scale integration: the evolution of design rules for biological automation. in *Annu Rev Biophys Biomol Struct*.36:213-31
33. Park S, Zhang Y, Lin S, Wang TH, & Yang S (2011) Advances in microfluidic PCR for point-of-care infectious disease diagnostics. *Biotechnology advances* 29(6):830-839.

34. Shi W, Qin J, Ye N, & Lin B (2008) Droplet-based microfluidic system for individual Caenorhabditis elegans assay. *Lab on a chip* 8(9):1432-1435.
35. Golberg A*, et al.* (2014) Cloud-enabled microscopy and droplet microfluidic platform for specific detection of Escherichia coli in water. *PloS one* 9(1):e86341.
36. Galas J-C, Bartolo D, & Studer V (2009) Active connectors for microfluidic drops on demand. *New Journal of Physics* 11(7):11.
37. Link DR*, et al.* (2006) Electric control of droplets in microfluidic devices. *Angewandte Chemie* 45(16):2556-2560.
38. Franke T, Abate AR, Weitz DA, & Wixforth A (2009) Surface acoustic wave (SAW) directed droplet flow in microfluidics for PDMS devices. *Lab on a chip* 9(18):2625-2627.
39. Gunther A & Jensen KF (2006) Multiphase microfluidics: from flow characteristics to chemical and materials synthesis. *Lab on a chip* 6(12):1487-1503.
40. Thorsen T, Roberts RW, Arnold FH, & Quake SR (2001) Dynamic pattern formation in a vesicle-generating microfluidic device. *Physical review letters* 86(18):4163-4166.
41. de Jong J*, et al.* (2006) Entrapped air bubbles in piezo-driven inkjet printing: Their effect on the droplet velocity. *Phys Fluids* 18(12).
42. Jahanshahi-Anbuhi S*, et al.* (2014) Pullulan encapsulation of labile biomolecules to give stable bioassay tablets. *Angewandte Chemie* 53(24):6155-6158.
43. Xia Y, Rogers JA, Paul KE, & Whitesides GM (1999) Unconventional Methods for Fabricating and Patterning Nanostructures. *Chemical reviews* 99(7):1823-1848.

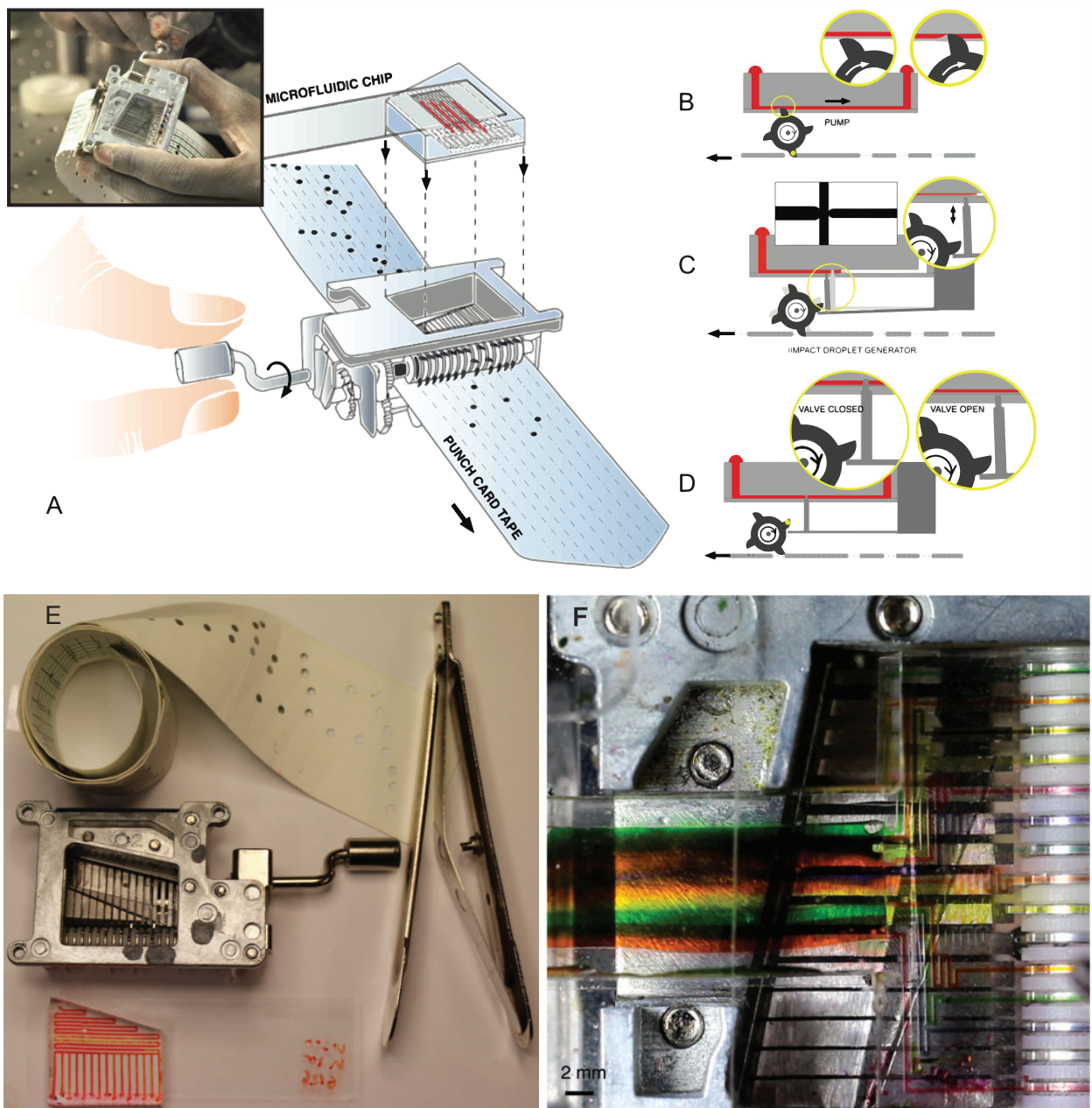

Figure 1. Punch card programmable microfluidic system comprised of punch card tape, microfluidic chip and actuator/driver (A); the inset in figure 1A is the photograph of the device in action. Pumping is achieved through the action of the other gear teeth that collapses the microchannel (B) and translates this pinch point along the channel. (C) On-demand droplet generation using an impact-based mechanism for a static droplet generator. (D) ON/OFF microfluidic valves based on a cantilever pin that is pushed against the chip. (E) The three separate components that make up our system and a hole-puncher for encoding punch card tape. (F) Top-down view of our device with 15 active channels.

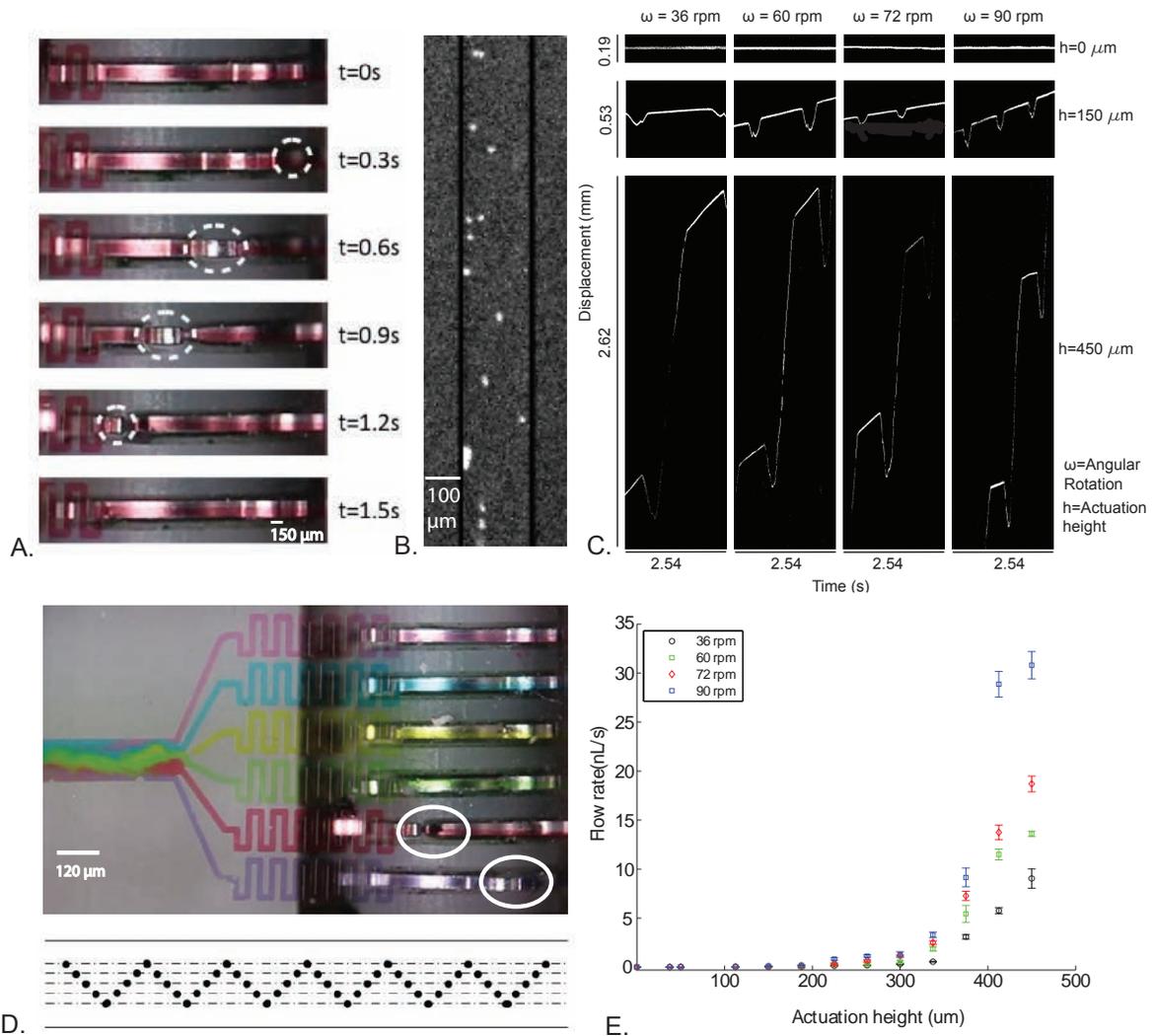

**Figure 2. Punch card controlled integrated multiplexed fluid pumps** (A) A series of images of a single channel coupled to the rotating gear disc during one pumping cycle. To characterize the flow pattern with each actuation, silver-coated glass beads were used in de-ionized water (B) and pumped through an open channel device. (C) Pumping in each cycle revealed a pulsatile oscillatory flow depicted in kymographs. The amplitude of directed unidirectional flow depends on actuation height (h) and the angular rotation (ω) from the hand-crank. (D) Top-down view of the microfluidic chip with simultaneous operation of six integrated pumps operated by a punch card tape with a zig-zag pattern. Fig. 2 E is the data for net positive flow as a function of h and ω.

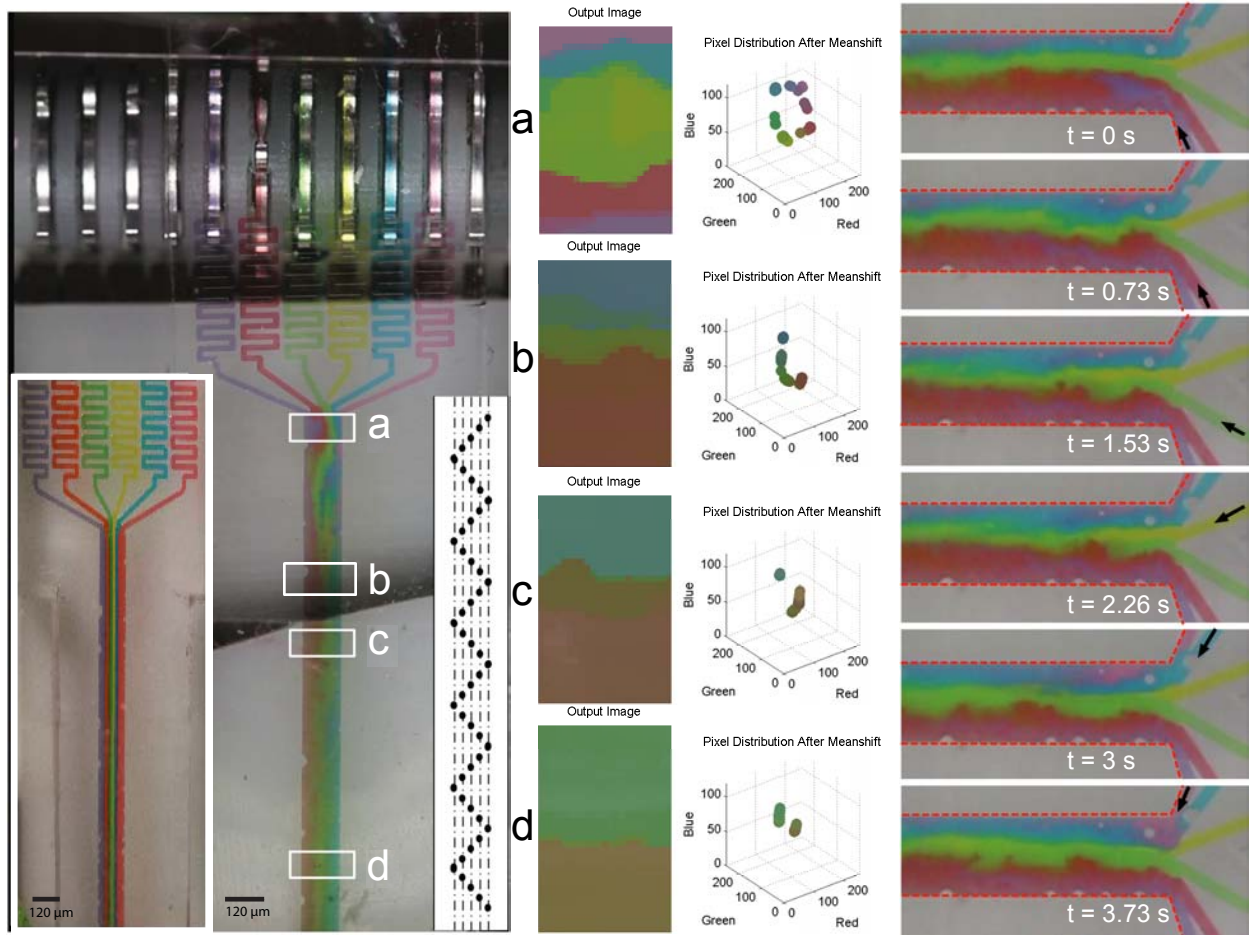

**Figure 3. Enhanced mixing is achieved using a zig-zag punch card code.** The figure shows a top-down image of six fluid channels pumped using our device. The left inset depicts fluids pumped through the device at a flow rate of ~100 nL/s using a syringe pump. The zig-zag pattern stretches interfaces and thus induces mixing in a 200 μm wide channel. Mixing is quantified by mean-shift cluster plots in four regions of the channel marked a,b,c,d along the outflow channel. Zoomed in image of region (a) reveals the pulsatile nature of flow that induces increased folding of neighboring fluid layers and thus enhancing diffusion and mixing.

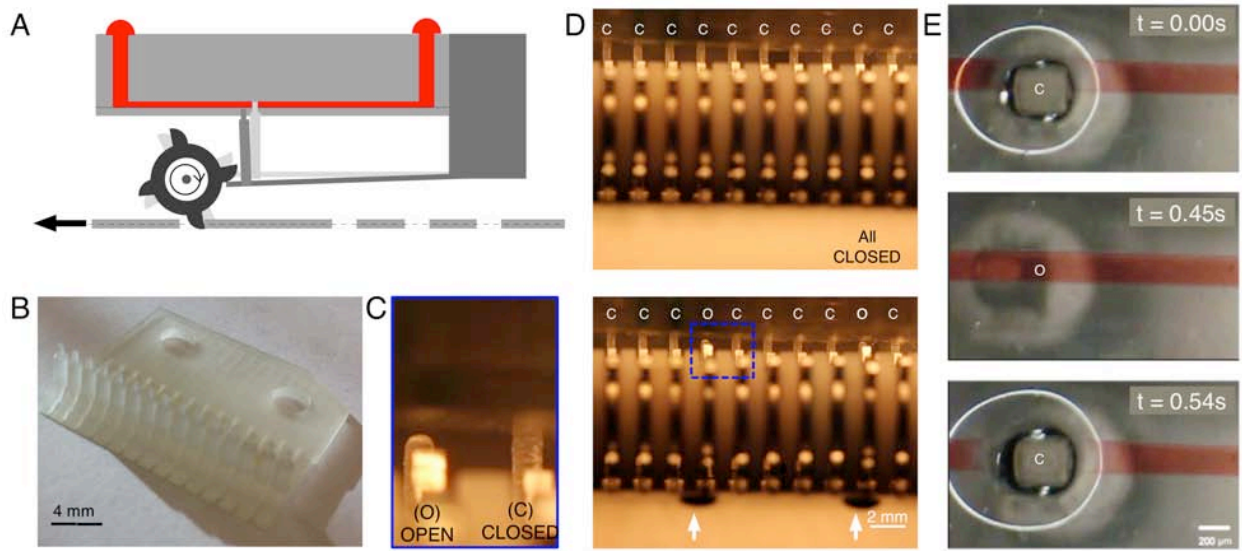

**Figure 4. Integrated punch card controlled ON/OFF valves.** (A) Schematic of valves showing coupling to the punch card code and the microfluidic chip. Image (B) is of a 3D printed cantilever beam with spaced pins 2mm apart, utilized for fifteen independent valves implemented. (C and D) are pictures of 15 independent valves implemented (side view), all of which could be independently actuated using the punch card tape. (E) Top view microphotograph of the valve in action, at a single instance of opening and closing showing a time of a complete ON/OFF cycle of 0.54 s with a single actuation. The image depicts the deformation of the completely collapsed channel.

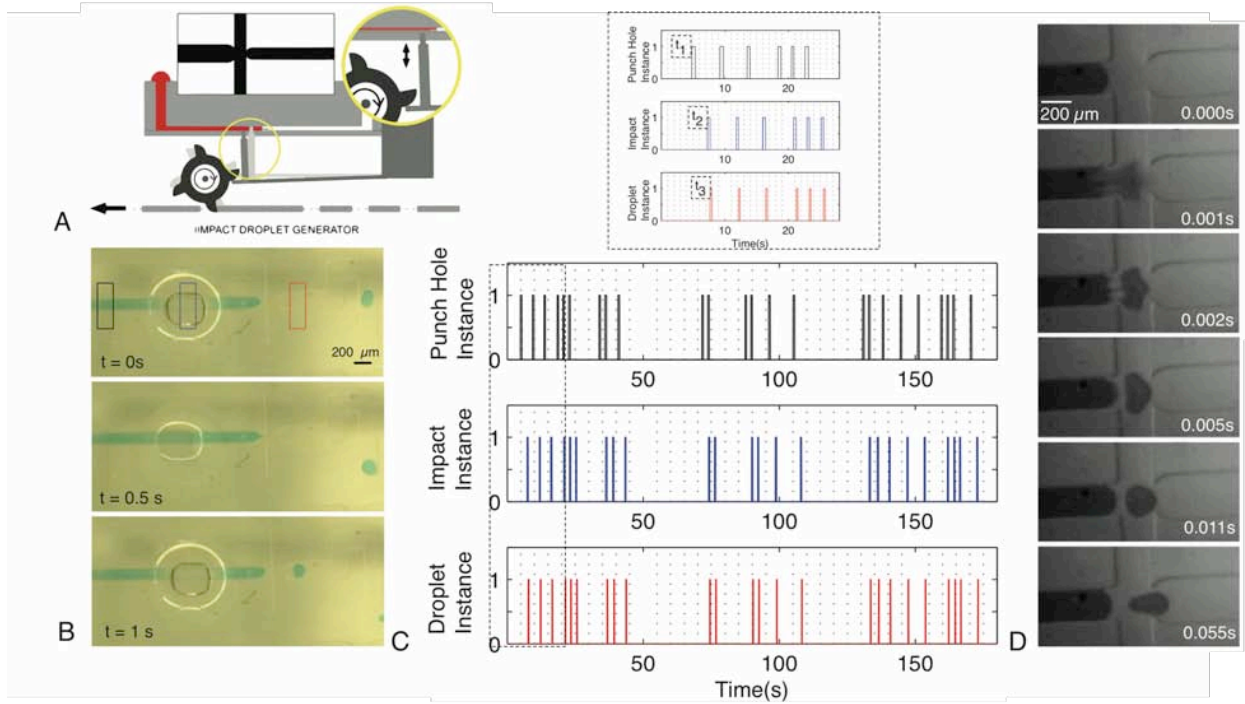

**Figure 5. Demonstration of impact based on-demand droplet generator** (A) Schematic of droplet generator showing plucking of the cantilever by a gear tooth leading to a pin impacting the channel, with each hole punched on the paper driving the actuator. (B) The impact droplet generator with a capillary pin valve flow-focusing geometry at three different time-points. The green fluid is food coloring in deionized water. Mineral oil was used as the carrier fluid. (C) Timing between the punched hole instance, the corresponding impact actuator action and the resultant droplet generation (inset depicts time delay between the three events). (D) High-speed imaging of droplet generation reveals an impact jet that forms in the first 2 ms of the impact. The jet is destabilized with formation of a narrow thread that breaks into a single droplet. Sequence of images depicts the entire operation over a period of 55 ms.

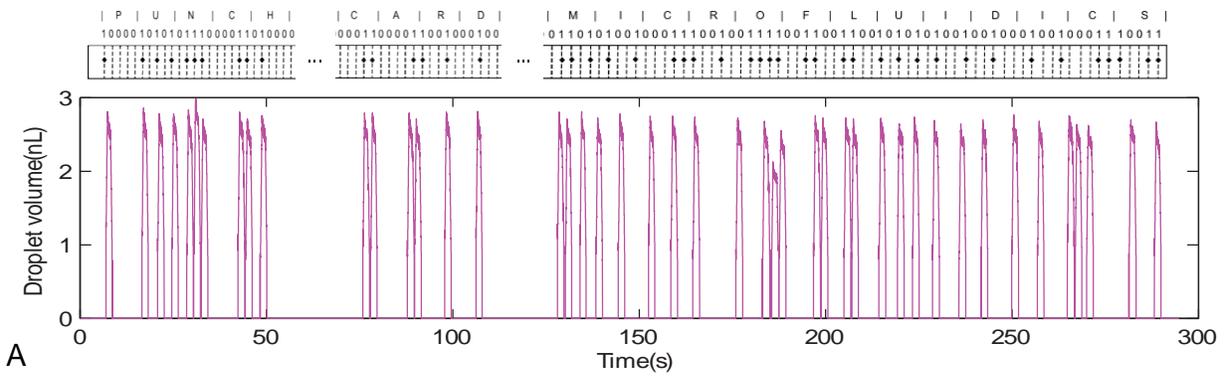
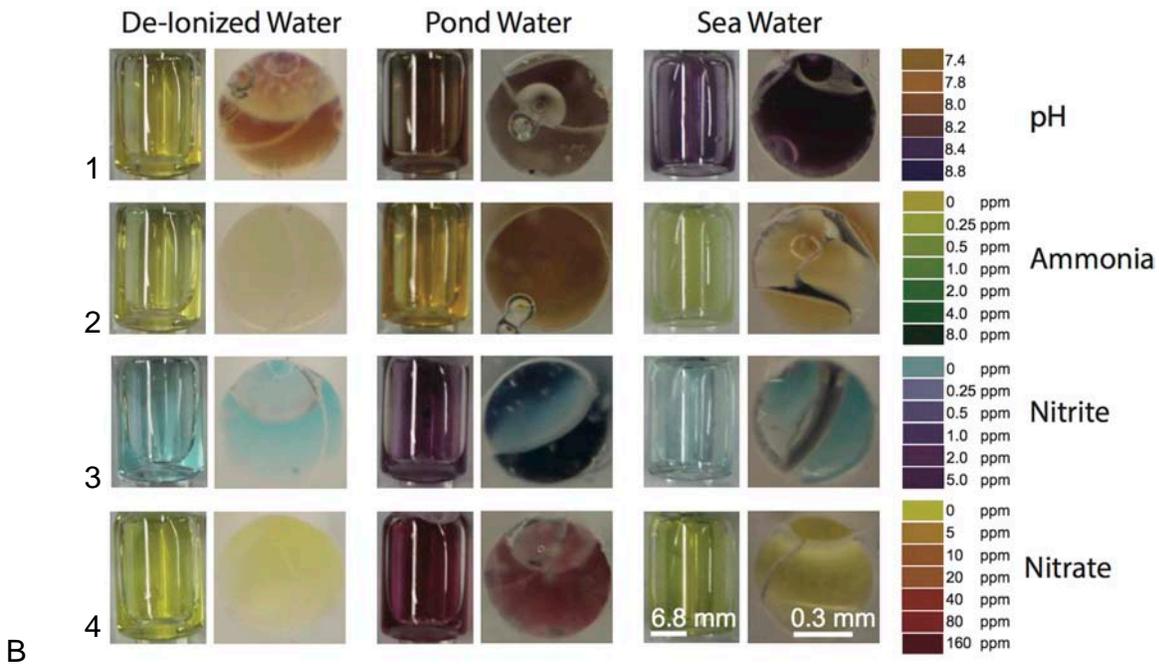

**Figure 6. One-to-one mapping and water testing application.** (A) One-to-one mapping of the code was demonstrated by implementing a 22 character binary code spelling "PUNCHCARD MICROFLUIDICS". The presence of a hole in punch card tape (top) results in a single droplet representing a binary value of "1" while the absence of a hole results in no droplet representing a binary value of "0". Vertical height of each plug represents droplet volume, which is approximately 2.5 nL. (B) Results of a qualitative colorimetric assay to test pH, ammonia, nitrates and nitrites in three samples: deionized water, pond water and seawater. The images are of the bulk sample (left image) and the outlet of our device (right image) with (1) pH readings with deionized water having a pH of 7.8, pond water at 8.2 and seawater at 8.8 (2) Ammonia levels in parts per million (or milligrams per deciliters) with deionized water having none, pond water having 0.25 and seawater having none as well (3) Nitrite levels with the deionized and sea water samples showing none and pond water having 2 ppm (4) Nitrate levels in the water samples with pond water having 160 ppm and the deionized water and sea water samples having none.

**Supplemental Information**

*Microfluidic chip design and fabrication*
The device was composed of two PDMS layers with different material properties bonded together. The first layer made of PDMS (polymer base to cross-linker ratio of 20:1) interfaced with the actuator and had the microfluidic channels. The second layer acted as an optically transparent firm-substrate against which the channels would be actuated while also being thick enough to ensure that the actuators interact with the micro channels effectively. This second layer was also made using a PDMS pre-polymer, mixed at a base to cross-linker ratio of 5:1. The layer with the channels measured approximately 500 µm while the thick layer measured approximately 7 mm). The PDMS mold for the microfluidic channels was made using SU-8 and a negative photoresist mask.

The fabricated mold was used to form the channels using PDMS by casting the pre-polymer that had a base to cross-linker ratio of 20:1. In order to enhance the actuators' effect on the microfluidic channels, the PDMS was cast by spin coating the layer of pre-polymer on the mold at 250 rpm for 30 s. The PDMS was then cured in an oven at 80°C for 30 minutes to make the layer that would be used sacrificially for the making of inlet and outlet holes punched through the second layer. This second layer with the base to cross-linker ratio of 5:1 had no features and so the pre-polymer mix was poured in a petri-dish and cured in an oven at 80°C for 30 minutes. When the time elapsed, the layer with channel features was peeled and placed on top of the thick layer. Inlet and outlet holes were then punched through both layers. Another casting iteration for the layer with the microfluidic channels was then done with pre-polymer being cured in an oven at 80°C for 30 minutes and used to replace the one that that was used sacrificially for the making of inlet and outlet holes. The two layers were then cured in an oven at 80°C for a minimum of four hours to ensure that they completely cured and bonded to each other. The cured and bonded PDMS device was then cut to shape and a glass slide was plasma bonded on the side that did not interface with the actuator. The glass slide only served as a handle for characterization purposes where the microfluidic device was raised or lowered as desired.

*Water quality test*
The test kit had reagents for testing pH from 6-7.6, high-range pH from 7.4 – 8.8, ammonia (up to 8 mg/dL), nitrites (up to 5 mg/dL) and nitrates (up to 160 mg/dL). The volumetric ratio of reagents to samples varied with the compound being tested.
A. Bulk solution test protocol
*i. pH bulk solution test protocol:* For the pH test, 5 ml of the bulk sample solution was mixed with 150 µl of the reagent solution. The solution was then left to sit in room temperature for 5 minutes before the colorimetric reading was taken. The color observed was compared against a color chart that corresponded to various pH values (Figure 5).
*ii. Ammonia bulk solution test protocol:* The ammonia test entailed the use of two reagents in 5 ml of the bulk sample solution. 240 µl of each of the two solutions was added to 5 ml of the sample and mixed together. The mixture was then left to sit in room temperature for five minutes before a colorimetric reading was done by comparing the color of the result against a chart with different ammonia concentrations (Figure 5).
*iii. Nitrite bulk solution test protocol.* In the nitrite test, 150 µl of the reagent solution was added to 5 ml of the sample to be tested and mixed. The resulting mixture was then left at room temperature for 5 minutes before the reading was made.

*Nitrate bulk solution test protocol:* The nitrate test required two reagent solutions to be added to the sample to be tested. 300 µl of each of the two reagents were added to 5ml of the sample and mixed together. The colorimetric result of the resulting solution was then read five minutes later.

*B. Microfluidic chip implementation protocol*

Each reagent test was carried out in a separate water line each having a volume 1.125 µl. For testing water pH, three fluid plugs each being 15 nL of fluid, generated by the punch card programmable platform were introduced into the water line. Five fluid plugs of the reagent were used for the high-range pH test and the nitrite test. Here a plug is defined as an actuation instance where the reagent is injected into the water solution. Each instance arises from a single punched hole. This translates to a volume of approximately 15 nL. The Ammonia test required 16 fluid plugs while the Nitrate test required 20 fluid plugs of the reagent being introduced into the sample water to be tested. The solutions were then incubated for five minutes before the color readout was done.

*Flow Velocity Characterization Experiments – Pumping*

The solution containing red fluorescent 2 µm monodisperse polysterene microbeads (Polysciences Catalog No. 18660-5) in deionized water at a concentration of approximately 10,000 beads per microliter was placed in a reservoir made using a 3 ml syringe devoid of the plunger. The fluid was introduced into the device using a 25 gauge blunt tip needle, flexible plastic tubing (Tygon, inner diameter 0.02 inches, Catalog no. AAQ04103) and a 25 gauge blunt tip that interfaced with the inlet hole (the same arrangement was used at the outlet leading to a waste chamber).

Figures S1

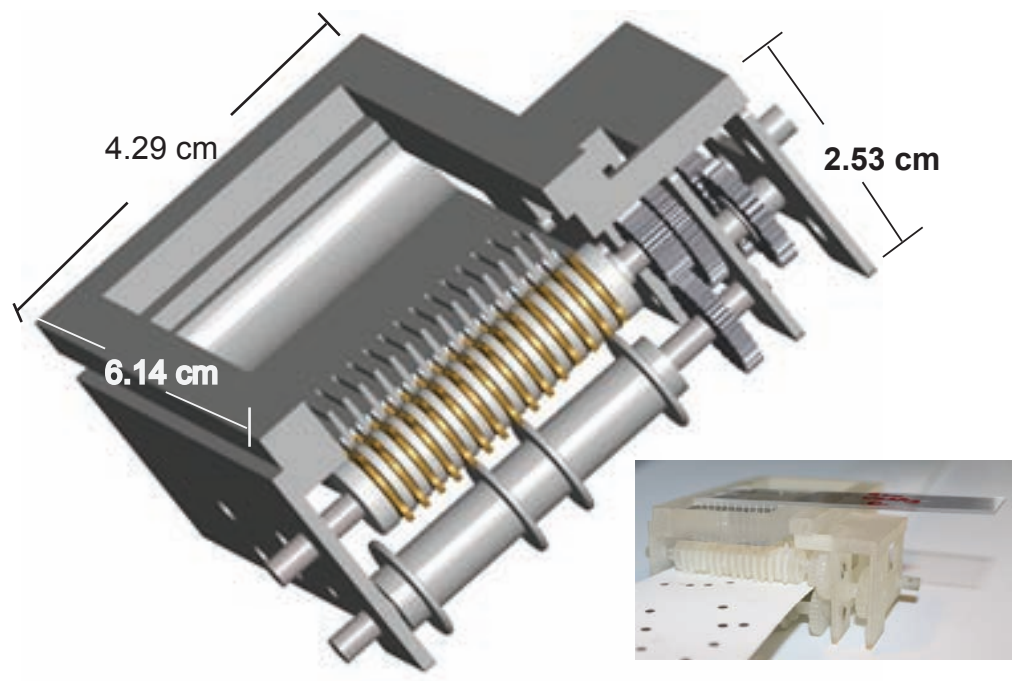

*Figure S1 A CAD rendition of the actuator/driver and functioning 3D-printed version of our device. The figure inset shows the coupling of both the punch card tape and the PDMS chip to the 3D-printed actuator The CAD model shows the actuator dimensions in cm.*

Movies

Movie S1. Valve pins

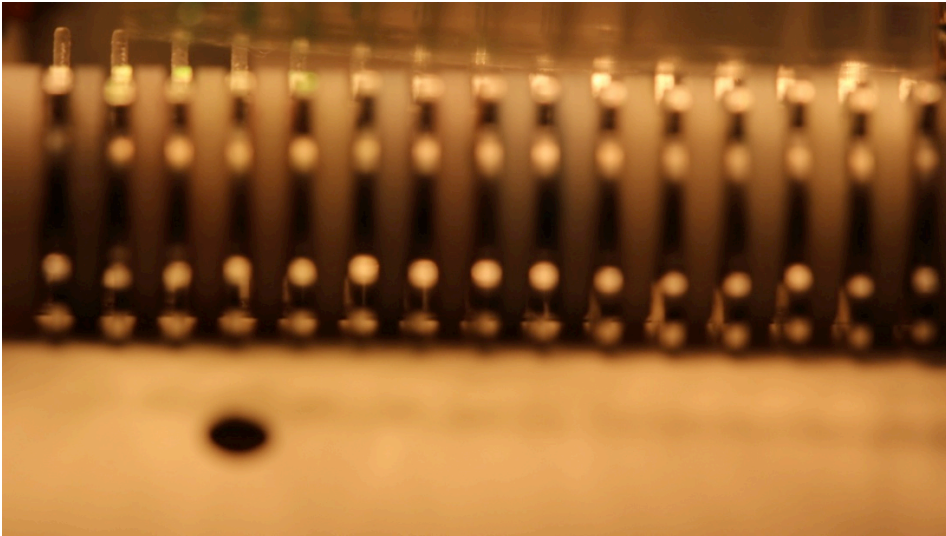

*Movie S1. The front view of the device in action wherein the punch card tape reels in towards the observer upon actuation. The image focus is on the valve pins that get actuated whenever there is a punched hole in the card. Presence of the holes results in the plucking action of the gear teeth on the cantilever beam that holds the pins leading to a forward and downward motion of the valve pins. The net downward movement results in the opening of the valve for the duration of the actuation (which is approximately 0.5 seconds). The capture and playback frame rate for the movie is 60 frames per second.*

Movie S2. Mixing

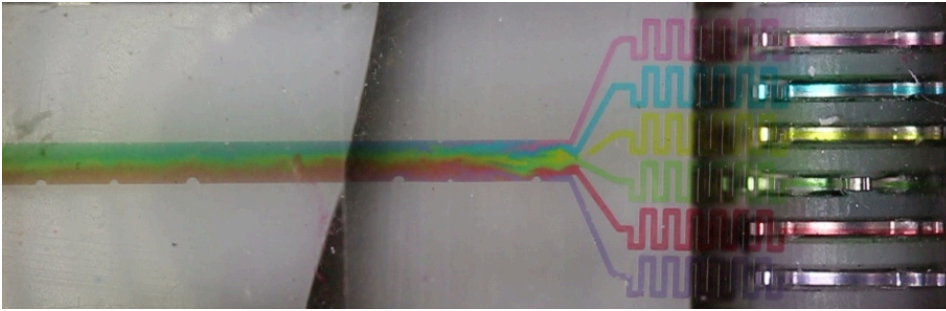

*Movie S2. The movie shown here is of the top-down view of the device in action. Six input channels, each having different food coloring solution are directed to an open chamber where mixing occurs. The pulsatile nature of the actuation results in enhanced mixing compared to flow generated using a syringe pump as a result of increased folding of fluid layers. The capture and playback frame rate for the movie is 60 frames per second.*

Movie S3. One-to-one mapping of droplet impact

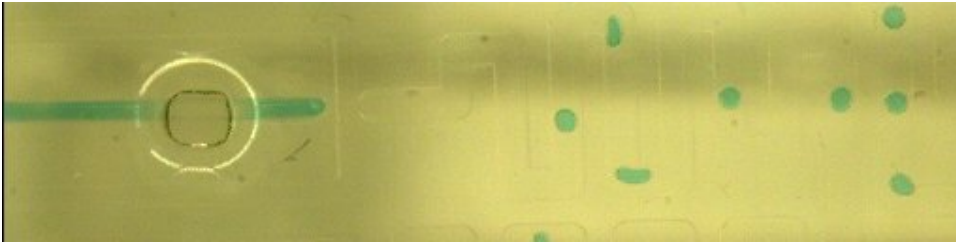

*Movie S3 The movie shown here is of the top-down view of our impact droplet generator. The actuating pin of the impact droplet generation mechanism formed a circular indentation on the PDMS upon complete closure of the channel. Mineral oil is used as a carrier fluid in a flow-focusing geometry. The droplets are of a solution composed of deionized water and food coloring. A capillary pin was implemented on the fluid line carrying the water-based solution, at the point of intersection with the mineral oil lines. With each punched hole, the actuator pin has a net downward force before springing back to its original position. The impact during this process results in the ejection of a fluid stream that forms a droplet and is carried downstream in the oil line. The capture and playback frame rate for the movie is 50 frames per second.*

Movie S4. High speed droplet generation

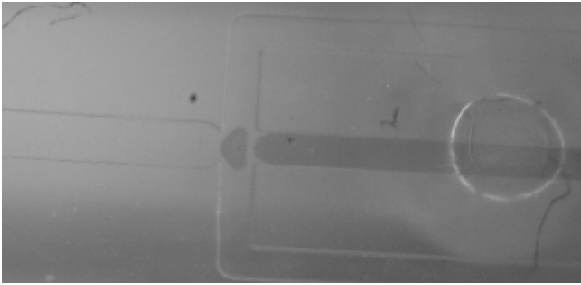

*Movie S4. The movie shown here is of a top down view of the device in action, to capture the sequence of events surrounding droplet generation using the impact-based technique. The movie was recorded at 1,000 frames per second. The circular indentation is of the actuator pin. As was the case in Movie S3, a flow-focusing geometry was implemented. However, in this case, unlike in Movie S3, a capillary valve was not implemented on the channel carrying the water-based solution, at the point where it intersects with the oil line. Mineral oil was used as the carrier fluid and deionized water with food-color used to generate the droplets in the carrier fluid. The capture frame rate for the movie is 1000 frames per second. The playback frame rate is 5 frames per second.*